\documentclass[preprint,onecolumn,aps,pre,eqsecnum,longbibliography,showpacs,showkeys,a4paper]{revtex4-1}
\usepackage[latin9]{inputenc}
\usepackage{color}
\definecolor{note_fontcolor}{rgb}{0.80078125, 0.80078125, 0.80078125}
\usepackage[english]{babel}
\usepackage{verbatim}
\usepackage{amsmath}
\usepackage{amssymb}
\usepackage{graphicx}
\usepackage[unicode=true,
 bookmarks=true,bookmarksnumbered=true,bookmarksopen=true,bookmarksopenlevel=1,
 breaklinks=true,pdfborder={0 0 0},backref=false,colorlinks=true]
 {hyperref}
\hypersetup{pdftitle={The role of current an spin},
 pdfauthor={Johannes Mesa Pascasio}}

\makeatletter

\providecommand{\tabularnewline}{\\}
\newenvironment{lyxgreyedout}
  {\textcolor{note_fontcolor}\bgroup\ignorespaces}
  {\ignorespacesafterend\egroup}

\@ifundefined{textcolor}{}
{%
 \definecolor{BLACK}{gray}{0}
 \definecolor{WHITE}{gray}{1}
 \definecolor{RED}{rgb}{1,0,0}
 \definecolor{GREEN}{rgb}{0,1,0}
 \definecolor{BLUE}{rgb}{0,0,1}
 \definecolor{CYAN}{cmyk}{1,0,0,0}
 \definecolor{MAGENTA}{cmyk}{0,1,0,0}
 \definecolor{YELLOW}{cmyk}{0,0,1,0}
 }
\numberwithin{equation}{section}
\numberwithin{figure}{section}
\numberwithin{table}{section}

\usepackage{graphicx}

\DeclareGraphicsExtensions{.png,.pdf,.eps}
\graphicspath{{pictures/}}

\makeatother

\begin{document}

\title{Born's Rule as Signature of a Superclassical Current Algebra}

\author{Siegfried \surname{Fussy}\textsuperscript{1}}

\email[E-mail: ]{ains@chello.at}

\homepage[Visit: ]{http://www.nonlinearstudies.at/}

\author{Johannes \surname{Mesa Pascasio}\textsuperscript{1,2}}

\email[E-mail: ]{ains@chello.at}

\homepage[Visit: ]{http://www.nonlinearstudies.at/}

\author{Herbert \surname{Schwabl}\textsuperscript{1}}

\email[E-mail: ]{ains@chello.at}

\homepage[Visit: ]{http://www.nonlinearstudies.at/}

\author{Gerhard \surname{Grössing}\textsuperscript{1}}

\email[E-mail: ]{ains@chello.at}

\homepage[Visit: ]{http://www.nonlinearstudies.at/}

\affiliation{\textsuperscript{1}Austrian Institute for Nonlinear Studies, Akademiehof\\
 Friedrichstr.~10, 1010 Vienna, Austria}

\affiliation{\textsuperscript{2}Institute for Atomic and Subatomic Physics, Vienna
University of Technology\\
Operng.~9, 1040 Vienna, Austria}

\affiliation{\vspace*{0cm}
}

\date{\today}
\begin{abstract}
We present a new tool for calculating the interference patterns and
particle trajectories of a double-, three- and \textit{N}-slit system
on the basis of an emergent sub-quantum theory developed by our group
throughout the last years. The quantum itself is considered as an
emergent system representing an off-equilibrium steady state oscillation
maintained by a constant throughput of energy provided by a classical
zero-point energy field. We introduce the concept of a ``relational
causality'' which allows for evaluating structural interdependences
of different systems levels, i.e.\ in our case of the relations between
partial and total probability density currents, respectively. Combined
with the application of 21\textsuperscript{st} century classical
physics like, e.g., modern nonequilibrium thermodynamics, we thus
arrive at a ``superclassical'' theory. Within this framework, the
proposed current algebra directly leads to a new formulation of the
guiding equation which is equivalent to the original one of the de\,Broglie-Bohm
theory. By proving the absence of third order interferences in three-path
systems it is shown that Born's rule is a natural consequence of our
theory. Considering the series of one-, double-, or, generally, of
\textit{N}-slit systems, with the first appearance of an interference
term in the double slit case, we can explain the violation of Sorkin's
first order sum rule, just as the validity of all higher order sum
rules. Moreover, the Talbot patterns and Talbot distance for an arbitrary
\textit{N}-slit device can be reproduced exactly by our model without
any quantum physics tool.%
\begin{lyxgreyedout}
\noindent \global\long\def\VEC#1{\mathbf{#1}}
\global\long\def\d{\,\mathrm{d}}
\global\long\def\e{{\rm e}}
\global\long\def\meant#1{\left<#1\right>}
\global\long\def\meanx#1{\overline{#1}}
\global\long\def\p{\partial}
\end{lyxgreyedout}

\end{abstract}


\maketitle

\section{Introduction}

In 1926, Born \cite{Born.1926quantenmechanik} suggested that \textbar{}$\psi(x,t)|^{2}$
is the probability to find the particle in the time interval $[t,t+\d t]$,
and in the length interval $[x,x+\d x]$. For different mutually excluding
paths of particles between source and detector one has to sum up the
$\psi$ functions of these paths coherently and then take the absolute
square of the linearly summed contributions. As a direct consequence
of this construction, a term appears describing the interference pattern
in the double slit diffraction experiment. Born's rule is one of the
key laws in quantum mechanics and it proposes that interference occurs
in pairs of possibilities, but never in triples etc. So-called multipath
interference terms representing interferences of higher order are
ruled out, be it in standard quantum mechanics or in the de\,Broglie-Bohm
theory, for example. Consequently, an addition of slits or paths does
\textit{not }increase the complexity of the whole system, but has
to be considered only quantitatively.

Although one can conclude that Born's rule is, at least indirectly,
confirmed by practically all quantum mechanical experiments within
the last hundred years, there had been no \textit{explicit} experiment
to support this proposition until a few years ago. The experimental
results hitherto seem to confirm the exact validity of Born's rule
up to the order of $10^{-4}$ \cite{Sinha.2009testing,Sinha.2010ruling,Sollner.2012testing}.
However, these experiments were commented critically by De\,Raedt
\textit{et\,al.}~\cite{DeRaedt.2012analysis}. The decomposition
of a three-path wave function into its lower order interference terms
might not correctly represent the experimental setup. So, it is still
an open question whether or not the mathematically correct derived
double slit contributions to the three slit result can be identified
with the sum of the experimentally derived double and single slit
contributions.

From the theoretical point of view no generally accepted derivation
of Born's rule has been given to date \cite{Landsman.2009born}, but
this does not imply that such a derivation is impossible in principle. 

In the following we try to shed light on this puzzle by combining
results of recently developed \textquotedblleft{}Emergent Quantum
Mechanics\textquotedblright{} \cite{Groessing.2012emerqum11-book}
with concepts of systems theory which we denote as ``relational causality''
\cite{Walleczek.2012mission}. Since the physics of different scales
is concerned, like, e.g., sub-quantum and classical macro physics,
we denote our sub-quantum theory as ``superclassical''. We consider
the quantum itself as an emergent system understood as off-equilibrium
steady state oscillation maintained by a constant throughput of energy
provided by the (\textquotedblleft{}classical\textquotedblleft{})
zero-point energy field. Starting with this concept, our group was
able to assess phenomena of standard quantum mechanics like Gaussian
dispersion of wave packets, superposition, double slit interference,
Planck's energy relation, or the Schrödinger equation, respectively,
as the emergent property of an underlying sub-structure of the vacuum
combined with diffusion processes reflecting the stochastic parts
of the zero-point field.

In Section~\ref{sec:path} we contrast the well-known physics behind
the double slit with an emergent vector field representation of the
observed interference field. In Section~\ref{sec:current} the essential
parts of our superclassical current algebra are presented and the
velocity field (corresponding to the guiding equation of the de\,Broglie-Bohm
theory) is derived. The crucial case testing the validity of Born's
rule by means of a three slit configuration is analyzed in Section~\ref{sec:three},
whereas the general $\mathit{N}$-slit setup is discussed in Section~\ref{sec:nslit}.
In Section~\ref{sec:conclusion} we summarize our results and give
an outlook on a possible breakdown of orthodox quantum mechanics representing
the emergent mean field theory out of our sub-quantum dynamics, consequently
associated with the violation of Born's rule.

\section{Interference and emergence at a Gaussian double slit\label{sec:path}}

Considering particles as oscillators (``bouncers'') coupling to
regular oscillations of the vacuum's zero-point field, which they
also generate, we have shown how a quantum can be understood as an
emergent system. In particular, the dynamics between the oscillator
and the ``bath'' of its thermal environment can be made responsible
not only for Gaussian diffraction at a single slit \cite{Groessing.2010emergence},
but also for the well-known interference effects at a double slit
\cite{Groessing.2012doubleslit}. We have shown that the quantum nature
of the spreading of the wave packet can be exactly described by combining
the convective with the orthogonal diffusive velocity fields. The
close resemblance of these two different velocities with the complex
velocity originally introduced by Schrödinger is discussed extensively
in \cite{Groessing.2010emergence}.

\begin{figure}[!tb]
\centering{}\includegraphics{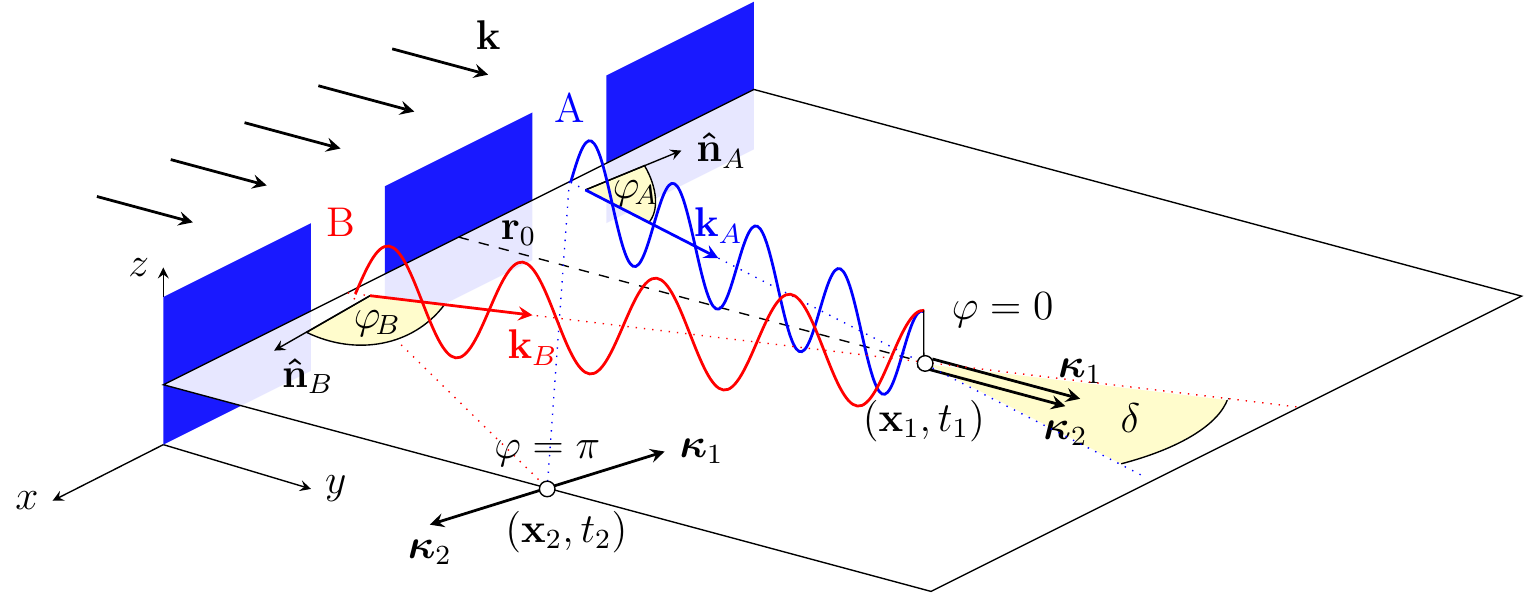}\caption{Geometry of interference at a double slit \textit{(A,B)} at exemplary
points $(\VEC x_{1},t_{1})$ and $(\VEC x_{2},t_{2})$.\label{fig:interf}}
\end{figure}

In Fig.~\ref{fig:interf} the underlying geometry for the wave vectors
is sketched, both for the classical interference and the emergent
case. For illustration, we show the two-dimensional setup where in
the classical picture the incoming wave vector $\VEC k=\frac{\mathrm{2\pi}}{\lambda}\VEC{\hat{k}}$
splits up at the Gaussian slits $\mathit{A}$ and $\mathit{B}$ into
$\VEC k_{A}$ and $\VEC k_{B}$, and both are orthogonal to the particular
propagating wave fronts. The respective phases for each of the beams
are usually denoted as $\varphi_{A(B)}=\VEC k_{A(B)}\cdot\VEC r_{0}$.
The phase difference $\varphi(\VEC x,t)$ reduces in case of coherent
plane waves to a time independent variable, i.e.\ $\varphi=\varphi_{A}-\varphi_{B}=\left(\VEC k_{A}-\VEC k_{B}\right)\cdot\VEC r_{0}$.
Depending on the size of the phase difference $\varphi\mathrm{\VEC{\mathrm{(}x\mathrm{)}}}$
at a specific point one obtains the well-known stationary interference
patterns, i.e.\ amplitude maxima in case of $\varphi=2n\pi$ and
minima for $\varphi=(2n+1)\pi$. Examples for both cases are shown
at the points $(\VEC x_{1},t_{1})$ and $(\VEC x_{2},t_{2})$, respectively.

We now take a closer look at the combined plane-wave amplitudes at
an arbitrary point $(\VEC x,t)$ of the spatio-temporal plane. The
amplitudes $R_{A(B)}=\sqrt{P_{A(B)}}$ associated with each of the
beams combine at $(\VEC x,t)$ to a total amplitude 
\begin{equation}
R=R_{A}\cos\varphi_{A}+R_{B}\cos\varphi_{B}.\label{eq:11a-1}
\end{equation}
By defining $\cos\varphi_{A(B)}=\VEC{\hat{n}_{\mathrm{A(B)}}\cdot\hat{k}_{\mathrm{A(B)}}}$
this can generally be written as 
\begin{equation}
R=\left(R_{A}\VEC{\hat{n}}_{A}\cdot\VEC{\hat{k}}_{A}+R_{B}\VEC{\hat{n}}_{B}\cdot\VEC{\hat{k}}_{B}\right).\label{eq:11b-2}
\end{equation}

In the emergent scenario we have to treat the two slits, or beam paths
respectively, as sources of two channels of a flow of probability
densities. The $N$ slits (or paths) are denoted in the following
as $A,\mathit{B,C,\ldots}$, whereas the emergent velocity channels
are characterized by the subscript $i=1,2,\ldots,N$, the reasoning
for which is given below. We look again at the -- in this case --
stationary phase difference $\varphi=\varphi_{1}-\varphi_{2}$, which
can also be understood as the arc cosine of two enclosed unit vectors
\begin{equation}
\varphi=\arccos(\VEC{\hat{s}}_{1}\cdot\VEC{\hat{s}}_{2})\,.\label{eq:phi}
\end{equation}
Again, in the stationary case $\VEC{\hat{s}}_{1}$ and $\VEC{\hat{s}}_{2}$
depend only on $\VEC x$ and represent two vector fields. 

According to our model, we interpret one possible solution of (\ref{eq:phi})
as the propagating wave vectors $\VEC{\boldsymbol{\kappa}}_{i}$ (also
termed ``convective velocity'' \cite{Sanz.2008trajectory}), which
both develop symmetrically to the axis given by $\VEC k_{A}+\VEC k_{B}$.
The emerging wave vectors $\VEC{\boldsymbol{\kappa}}_{i}$ relate
to the velocities via $\VEC p_{i}=\hbar\VEC{\boldsymbol{\kappa}}_{i}$
and thus $\VEC v_{i}=\frac{\hbar}{m}\VEC{\boldsymbol{\kappa}}_{i}$,
which generally do not coincide with the original directions of $\VEC k_{A(B)}$. 

Here some words of caution are appropriate. The vector fields $\VEC{\boldsymbol{\kappa}}_{i}$
have emergent properties, thus the index \textit{i }cannot be understood
as a direct link to the slits $\mathit{A}$ or $\mathit{B}$, respectively.
In fact, we can only state that the setup of the double slit gives
rise to two independent probability channels, which we denote with
the indices \textit{$i$.} Consequently, the enclosed angle $\varphi$
of the two emerging velocities $\VEC v_{i}$ should not be confused
with the geometric angle $\delta$ of the spreading waves. Note that
the differences between the incoming wave vectors $\VEC k_{A(B)}$
on one hand, and the emerging wave fields $\VEC{\boldsymbol{\kappa}}_{i}$
on the other, are in complete analogy to those between the geometric
rays and the streamlines in optical currents, as can be seen impressively
from Fig.~3 in \cite{Berry.2009optical}.

The second possible solution of (\ref{eq:phi}) are the orthogonal
osmotic velocities $\VEC u_{i}$, again for the two channels of probability
densities. Clearly, the original amplitudes $R_{A(B)}$ have to be
identified with those of the emergent vectors $R_{i}$, $i=1,2$.
The osmotic velocities $\VEC u_{i}$ refer to diffusion processes
reflecting the stochastic fluctuations of the zero-point field. For
the rest of this paper, we keep the notations $\VEC v_{i}$ and $\VEC u_{i}$
to denote the emergent convective and diffusive fields, respectively.

As can be seen from Fig.~\ref{fig:interf}, at $(\VEC x,t)$ the
new, emerging wave vector or velocity field, respectively, discloses
new information with regard to the amplitudes at that point: The phase
difference is nonlocally encoded at each point of the plane due to
the Gaussians \textit{not} being truncated, as it will be discussed
in the last section. Since the exemplary point $\VEC{\mathrm{(}x_{\mathrm{1}}},t_{1}$)
lies on the central symmetrical line between slit $\mathit{A}$ and
$\mathit{B}$, both paths from the slits are of equal length, with
the consequence of a vanishing phase difference. Therefore, both of
the emergent wave vectors $\VEC{\boldsymbol{\kappa}}_{1}$ and $\VEC{\boldsymbol{\kappa}}_{2}$
have to be parallel at $\VEC{\mathrm{(}x_{\mathrm{1}}},t_{1}$). Analogously,
in the case of the destructive interference at $\mathrm{(}\VEC x_{\mathrm{2}},t_{2})$,
$\VEC{\boldsymbol{\kappa}}_{\mathrm{1}}$ and $\VEC{\boldsymbol{\kappa}}_{\mathrm{2}}$
point into opposite directions.

The emergent diffusive velocities have to fulfill the condition of
being unbiased w.r.t.~the convective velocities, i.e.\ the orthogonality
relation for the \textit{averaged} velocities derived in \cite{Groessing.2010emergence}:
$\VEC{\overline{vu}}=0$, since any fluctuations $\VEC u=\delta\left(\nabla S/m\right)$
are shifts along the surfaces of action $\mathit{S=\mathrm{\mathrm{const}}.}$ 

\begin{figure}[!tb]
\centering{}\includegraphics{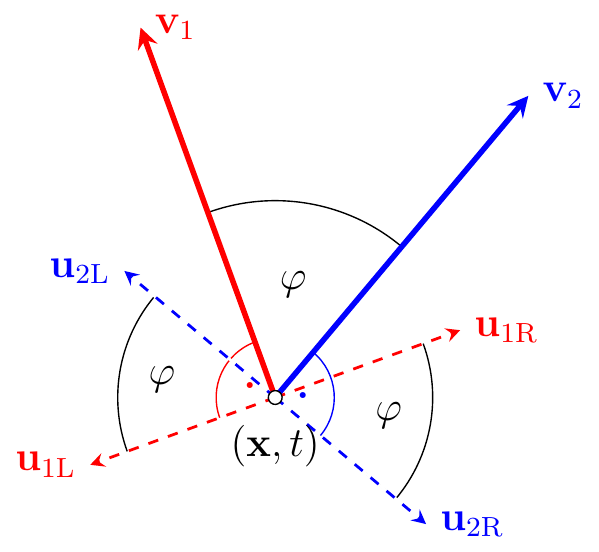}\caption{Geometry of emergent velocities and relative phases for a two-beam
setup.\label{fig:vektor}}
\end{figure}

Each point of the probability (or amplitude) landscape evolves on
the spatiotemporal plane according to the emergent propagation velocities
$\VEC v_{i}(\VEC x,t)$, $i=1,2$. In addition, the differential equations
for $\VEC u(\VEC x,t)$ describe the dispersion of the Gaussians and
split up into $\VEC u_{\mathrm{1}}(\VEC x,t)$ and $\VEC u_{\mathrm{2}}(\VEC x,t)$
for each channel. The enclosed phase difference angles $\varphi$
can be found between $\VEC v_{i}$ and the two opposed components,
of $\VEC u_{i}$, $i=1,2$, respectively, denoted as $\VEC u_{i\mathrm{R}}$
and $\VEC u_{i\mathrm{L}}$ (Fig.~\ref{fig:vektor}). Since $\VEC u_{i\mathrm{R(L)}}$
are orthogonal to $\VEC v_{\mathrm{i}}$, all other enclosed angles
like, e.g., the one between the unit vectors $\VEC{\hat{v}}_{1}$
and $\VEC{\hat{u}}_{2R}$, can be expressed in terms of $\varphi$,
i.e.\ $\sphericalangle(\VEC{\hat{v}}_{1},\VEC{\hat{u}}_{2\mathrm{R}})=\frac{\pi}{2}+\varphi$,
for example.

In the following we will see how the trajectories which represent
averaged paths due to the averaged velocities, can be calculated with
the help of a ``superclassical current'' algebra leading to the
expressions for the total current $\VEC{J_{\mathrm{tot}}}$ and the
total probability density $P_{\mathrm{tot}}$ at $(\VEC x,t)$. 

To account for the different velocity channels $i=1,\ldots,3N$, $N$
being the number of slits, we now introduce for general cases generalized
velocity vectors $\VEC w_{i}$, with 
\begin{equation}
\VEC w_{1}:=\VEC v_{\mathrm{1}},\quad\VEC w_{2}:=\VEC u_{\mathrm{1R}},\quad\VEC w_{3}:=\VEC u_{\mathrm{1L}}\label{eq:nslit.2.4}
\end{equation}
for the first channel, and 
\begin{equation}
\VEC w_{4}:=\VEC v_{\mathrm{2}},\quad\VEC w_{5}:=\VEC u_{\mathrm{2R}},\quad\VEC w_{6}:=\VEC u_{\mathrm{2L}}\label{eq:nslit.2.5}
\end{equation}
for the second channel in the case of $N=2$. The associated amplitudes
$R(\VEC w_{i})$ for each channel are taken to be the same, i.e.\ $R(\VEC w_{1})=R(\VEC w_{2})=R(\VEC w_{3})=R_{1}$,
and $R(\VEC w_{4})=R(\VEC w_{5})=R(\VEC w_{6})=R_{2}$. This renumbering
procedure will turn out as an important practical bookkeeping tool.

\section{A superclassical current algebra\label{sec:current}}

Generally, a probability density current is defined as $\VEC J=P\VEC v$.
To calculate the total average current, we sum up all contributions
of said vectors $\VEC v_{i}(\VEC x,t)$ and $\VEC u_{i\mathrm{R}(L)}(\VEC x,t)$
for each point $(\VEC x,t)$ of the plane by matching the unit velocity
vector component, associated with each vector and multiplied with
the corresponding amplitude $R(\VEC w_{i})$, one by one with all
other unit vector components together with their amplitudes. In other
words, the corresponding probability density $P(\VEC w_{i})$ for
any channel or, respectively, for any velocity component $\VEC w_{i}$,
is obtained by the pairwise projection on the unit vector $\VEC{\hat{w}}_{i}$
weighted by $R(\VEC w_{i})$ of the totality of all amplitude weighted
unit velocity vectors being operative at $\mathrm{(}\VEC x,t)$. 

\begin{figure}[!tb]
\begin{centering}
\includegraphics{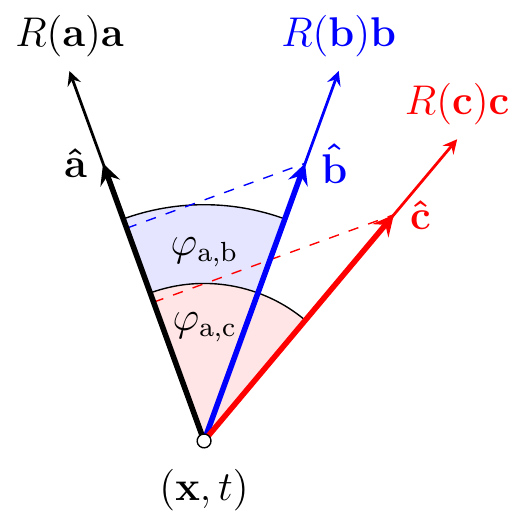}
\par\end{centering}

\caption{Scheme for the projection rule to obtain the projected contributions
from partial currents $\VEC{J\mathrm{(}b\mathrm{)}}$ and $\VEC{J\mathrm{(}c\mathrm{)}}$
for the partial current $\VEC{J\mathrm{(}a\mathrm{)}}$.\label{fig:Scheme} }
\end{figure}

In Fig.~\ref{fig:Scheme} the projection scheme generating the partial
current $\VEC{J\mathrm{(}a\mathrm{)}}$ is shown symbolically for
a total of three velocity channels $\VEC a$, $\VEC b,$ and $\VEC c$.
The projections of the unit vectors $\VEC{\hat{b}}$ and $\hat{\VEC c}$
of the second and third velocity vectors are indicated as dashed lines.
The probability density $P(\VEC a)$ for said current is built by
the products $P(\VEC a)=R(\VEC a)\VEC{\hat{a}}\cdot{\displaystyle [}\VEC{\hat{a}}R(\VEC a)+\VEC{\hat{b}}R(\VEC b)+\hat{\VEC c}R(\VEC c)]=R^{2}(\VEC a)+R(\VEC a)R(\VEC b)\cos\varphi_{\mathrm{a},\mathrm{b}}+R(\VEC a)R(\VEC c)\cos\varphi_{\mathrm{a,c}}$,
with $\VEC{\hat{a}}\cdot\VEC{\hat{b}}=\cos\varphi_{\mathrm{a},\mathrm{b}}$,
etc.

In case of only two velocities $\VEC a$ and $\VEC b$, one immediately
sees the resemblance with the classical interference amplitude of
Eq.~(\ref{eq:11a-1}): $P(\VEC a)+P(\VEC b)=R^{2}(\VEC a)+2R(\VEC a)R(\VEC b)\cos\varphi_{\mathrm{a},\mathrm{b}}+R^{2}(\VEC b)$,
with the main difference consisting in the phase difference as the
included angle between $\VEC{\hat{a}}$ and $\VEC{\hat{b}}$ according
to Eq.~(\ref{eq:phi}).

The two-path set-up has $3N=6$ velocity vectors at each point (cf.\ Figs.~\ref{fig:vektor}
and \ref{fig:Scheme} and Eqs.~(\ref{eq:nslit.2.4}) and (\ref{eq:nslit.2.5}))
and we obtain for the partial intensities and currents, i.e.\ for
each channel component $\mathit{i}$
\begin{align}
P(\VEC w_{i}) & =R(\VEC w_{i})\VEC{\hat{w}}_{i}\cdot{\displaystyle \sum_{j=1}^{6}}\VEC{\hat{w}}_{j}R(\VEC w_{j})\label{eq:Proj}\\
\VEC J\mathrm{(}\VEC w_{i}\mathrm{)} & =\VEC w_{i}P(\VEC w_{i}),\quad i=1,\ldots,6,
\end{align}
with
\begin{equation}
\cos\varphi_{i,j}:=\VEC{\hat{w}}_{i}\cdot\VEC{\hat{w}}_{j}.
\end{equation}
Consequently, the total intensity and current read as
\begin{align}
P_{\mathrm{tot}}= & {\displaystyle \sum_{i=1}^{6}}P(\VEC w_{i})=\left({\displaystyle \sum_{i=1}^{6}}\VEC{\hat{w}}_{i}R(\VEC w_{i})\right)^{2},\label{eq:Ptot6}\\
\VEC J_{\mathrm{tot}}= & \sum_{i=1}^{6}\VEC J(\VEC w_{i})={\displaystyle \sum_{i=1}^{6}}\VEC w_{i}P(\VEC w_{i}),\label{eq:Jtot6}
\end{align}
 leading to the \textit{emergent total velocity}

\begin{equation}
\VEC v_{\mathrm{tot}}=\frac{\VEC J_{\mathrm{tot}}}{P_{\mathrm{tot}}}=\frac{{\displaystyle \sum_{i=1}^{6}}\VEC w_{i}P(\VEC w_{i})}{{\displaystyle \sum_{i=1}^{6}}P(\VEC w_{i})}\,.\label{eq:vtot_fin}
\end{equation}

Thus we obtain phase-dependent amplitude contributions of the total
system's wave field projected on each channel's amplitude at point
$\mathrm{(}\VEC x,t)$ via the conditional probability $P(\VEC w_{i})$.
The local intensity of a partial current is dependent on all other
currents, and the total current itself is composed of all partial
components. This mutual dependence of a current's ``totality'' and
its parts, we denote as ``relational causality'' \cite{Walleczek.2012mission},
and this constitutes the essential part of what we call a ``superclassical''
current algebra. We denote a theory as \textquotedblleft{}superclassical\textquotedblright{}
if it covers a vast range of spatio-temporal scales, i.e.\ from a
usual classical down to a hypothesized sub-quantum domain. In assuming
the sub-quantum domain to be described in terms of modern classical
physics, quantum theory thus appears as sandwiched between two classical
regimes. \textquotedblleft{}Super-classical\textquotedblright{} is
introduced in close analogy to the use of the term \textquotedblleft{}superstatistics\textquotedblright{},
which itself not only relates vastly disparate space-time scales,
but also shows highly unexpected emergent behavior on intermediate
scales \cite{Jizba.2012emergence,Jizba.2012quantum}. We thus presume
that quantum phenomenology emerges from the superposition of processes
on said vastly disparate scales. The term ``current algebra'' is
borrowed from quantum field theory, but used here in a very broad
sense, with \char`\"{}algebra\char`\"{} referring to its original
meaning as \char`\"{}reunion of broken parts\char`\"{}, i.e.\ in
the sense of a proper combinatorics of currents. However, both of
said concepts share the property of using currents as \textit{basic}
ingredient and not as derivation of some elementary entity like, e.g.,
an elementary particle. 

Note that the usual symmetry (cf.\ the classical interference case
above) between $P(\VEC w_{i})$ and $R(\VEC w_{i})$ is broken: $P(\VEC w_{i})\neq R^{2}(\VEC w_{i})$,
i.e.\ although to each velocity component $\VEC w_{i}$ an amplitude
$R(\VEC w_{i})$ is associated, the \textit{partial} probability density
$P(\VEC w_{i})$ is not the mere squared amplitude any more. The consequences
of this asymmetry are discussed in the following section.

Returning now to our previous notation for the six velocity components
$\VEC v_{i}$, $\VEC u_{i\mathrm{R}}$, $\VEC u_{i\mathrm{L}}$, $i=1,2$,
the partial current associated with $\VEC v_{\mathrm{1}}$ originates
from building the scalar product of $\VEC{\hat{v}}_{1}$ with all
other unit vector components and reads as
\begin{equation}
\VEC J(\VEC v_{\mathrm{1}})=\VEC v_{\mathrm{1}}P(\VEC v_{\mathrm{1}})=\VEC v_{\mathrm{1}}R_{1}\VEC{\hat{v}}_{1}\cdot(\VEC{\hat{v}}_{1}R_{1}+\VEC{\hat{u}_{\mathrm{1R}}\mathrm{\mathit{R_{\mathrm{1}}}}+}\VEC{\hat{u}_{\mathrm{1L}}\mathrm{\mathit{R_{\mathrm{1}}}}+}\VEC{\hat{v}}_{2}R_{2}+\VEC{\hat{u}_{\mathrm{2R}}\mathit{R_{\mathrm{2}}}+}\VEC{\hat{u}}_{\mathrm{2L}}R_{2}).\label{eq:Jv1}
\end{equation}

Since trivially 
\begin{equation}
\VEC{\hat{u}}_{i\mathrm{R}}R_{i}+\VEC{\hat{u}}_{i\mathrm{L}}R_{i}=0,\quad i=1,2,\label{eq:triv}
\end{equation}
Eq.~(\ref{eq:Jv1}) leads to
\begin{align}
\VEC J\mathrm{(}\VEC v_{\mathrm{1}})=\VEC v_{\mathrm{1}}\left(R_{1}^{2}+R_{1}R_{2}\cos\varphi\right),
\end{align}
which results from the representation of the emerging velocity fields
according to Eq.~(\ref{eq:phi}), since we get the cosine of the
phase difference $\varphi$ as a natural result of the scalar product
of the velocity vectors $\VEC v_{i}$. The non-zero residua of the
other vector fields yield 
\begin{equation}
\VEC J\mathrm{(}\VEC u_{\mathrm{1\mathrm{R}}}\mathrm{)}=u_{\mathrm{1R}}P\mathrm{(}\VEC u_{\mathrm{1\mathrm{R}}}\mathrm{)}=\VEC u_{\mathrm{1R}}\left(R_{1}\VEC{\hat{u}}_{\mathrm{1R}}\cdot\VEC{\hat{v}}_{2}R_{2}\right)=\VEC u_{\mathrm{1R}}R_{1}R_{2}\cos\left(\frac{\pi}{2}-\varphi\right)=\VEC u_{\mathrm{1R}}R_{1}R_{2}\sin\varphi
\end{equation}
and 
\begin{equation}
\VEC J\mathrm{(}\VEC{\VEC u_{\mathrm{1L}}})=\VEC u_{\mathrm{1L}}P\mathrm{(}\VEC u_{\mathrm{1L}}\mathrm{)}=\VEC u_{\mathrm{1L}}\left(R_{1}\VEC{\hat{u}}_{\mathrm{1L}}\cdot\VEC{\hat{v}}_{2}R_{2}\right)=\VEC u_{\mathrm{1L}}R_{1}R_{2}\cos\left(\frac{\pi}{2}+\varphi\right)=-\VEC u_{\mathrm{1L}}R_{1}R_{2}\sin\varphi.
\end{equation}
Analogously, we obtain for the convective velocity vector field of
the second channel
\begin{equation}
\VEC{J\mathrm{(}v_{\mathit{\mathrm{2}}}\mathrm{)}}=\VEC v_{\mathrm{2}}P(\VEC v_{\mathrm{2}})=\VEC v_{2}\left(R_{2}^{2}+R_{1}R_{2}\cos\varphi\right).
\end{equation}
The corresponding diffusive velocity vector fields read as
\begin{align}
\VEC J\mathrm{(}\VEC u_{\mathrm{2R}})= & \VEC u_{2R}P\mathrm{(}\VEC u_{2\mathrm{R}}\mathrm{)}=\VEC u_{2R}\left(R_{2}\VEC{\hat{u}}_{\mathrm{2R}}\cdot\VEC{\hat{v}}_{1}R_{1}\right)=\VEC u_{2R}R_{1}R_{2}\cos\left(\frac{\pi}{2}+\varphi\right)=-\VEC u_{2R}R_{1}R_{2}\sin\varphi,\\
\VEC J\mathrm{(}\VEC u_{\mathrm{2L}})= & \VEC u_{\mathrm{2L}}P\mathrm{(}\VEC u_{2\mathrm{L}}\mathrm{)}=\VEC u_{\mathrm{2L}}\left(R_{2}\VEC{\hat{u}}_{\mathrm{2L}}\cdot\VEC{\hat{v}}_{1}R_{1}\right)=\VEC u_{\mathrm{2L}}R_{1}R_{2}\cos\left(\frac{\pi}{2}-\varphi\right)=\VEC u_{\mathrm{2L}}R_{1}R_{2}\sin\varphi.
\end{align}
Note that the nontrivial sine contributions to the total current stem
from the projections between the diffusive velocities $\VEC u_{\mathrm{1R(L)}}$
of the first channel on the unit vector $\VEC{\hat{v}}_{2}$ of the
convective velocity of the second channel, and \textit{vice versa}.
Combining all terms, we obtain the result for the total current
\begin{align}
\VEC J_{\mathrm{tot}} & =\VEC v_{\mathrm{1}}P(\VEC v_{\mathrm{1}})+\VEC u_{1R}P\mathrm{(}\VEC u_{1\mathrm{R}}\mathrm{)}+\VEC u_{1\mathrm{L}}P\mathrm{(}\VEC u_{\mathrm{1\mathrm{L}}}\mathrm{)}+\VEC v_{\mathrm{2}}P(\VEC v_{\mathrm{2}})+\VEC u_{2R}P\mathrm{(}\VEC u_{2\mathrm{R}}\mathrm{)}+\VEC u_{\mathrm{2L}}P\mathrm{(}\VEC u_{2\mathrm{L}}\mathrm{)}\nonumber \\
 & =R_{1}^{2}\VEC v_{\mathrm{1}}+R_{2}^{2}\VEC v_{\mathrm{2}}+R_{1}R_{2}\left(\VEC v_{\mathrm{1}}+\VEC v_{\mathrm{2}}\right)\cos\varphi+R_{1}R_{2}\left([\VEC u_{1\mathrm{R}}-\VEC u_{1\mathrm{L}}]-[\VEC u_{2\mathrm{R}}-\VEC u_{2\mathrm{L}}]\right)\sin\varphi.\label{eq:Jtot}
\end{align}
The resulting diffusive velocities $\VEC u_{i\mathrm{\mathrm{R}}}-\VEC u_{i\mathrm{L}}$
are identified with the effective diffusive velocities $\VEC u_{i}$
for each channel. Note that \textit{one} of those velocities, $\VEC u_{i\mathrm{\mathrm{R}}}$
or $\VEC u_{i\mathrm{L}}$, respectively, is always zero, so that
the product of said difference with $\sin\varphi$ guarantees the
correct sign of the last term in Eq.~(\ref{eq:Jtot}). Thus we obtain
the final expression for the total density current built from the
remaining $2N=4$ velocity components
\begin{equation}
\VEC J_{\mathrm{tot}}=R_{1}^{2}\VEC v_{\mathrm{1}}+R_{2}^{2}\VEC v_{\mathrm{2}}+R_{1}R_{2}\left(\VEC v_{\mathrm{1}}+\VEC v_{2}\right)\cos\varphi+R_{1}R_{2}\left(\VEC u_{1}-\VEC u_{2}\right)\sin\varphi.\label{eq:Jfinal}
\end{equation}
Summing up the probabilities associated with each of the partial currents
we get according to the ansatz (\ref{eq:Proj}) and the relations
(\ref{eq:Ptot6}) and (\ref{eq:triv})
\begin{align}
P_{\mathrm{tot}} & =(R_{1}\VEC{\hat{v}}_{1}+R_{1}\VEC{\hat{u}}_{\mathrm{1R}}+R_{1}\VEC{\hat{u}}_{\mathrm{1L}}+R_{2}\VEC{\hat{v}}_{2}+R_{2}\VEC{\hat{u}}_{\mathrm{2R}}+R_{2}\VEC{\hat{u}}_{\mathrm{2L}})^{2}\nonumber \\
 & =(R_{1}\VEC{\hat{v}}_{1}+R_{2}\VEC{\hat{v}}_{2})^{2}=R_{1}^{2}+R_{2}^{2}+2R_{1}R_{2}\cos\varphi=P(\VEC v_{\mathrm{1}})+P(\VEC v_{\mathrm{2}}).\label{eq:Ptot2slit}
\end{align}
The total velocity $\VEC{v_{\mathrm{tot}}}$ according to Eq.~(\ref{eq:vtot_fin})
now reads as
\begin{equation}
\VEC v_{\mathrm{tot}}=\frac{R_{1}^{2}\VEC v_{\mathrm{1}}+R_{2}^{2}\VEC v_{\mathrm{2}}+R_{1}R_{2}\left(\VEC v_{\mathrm{1}}+\VEC v_{2}\right)\cos\varphi+R_{1}R_{2}\left(\VEC u_{1}-\VEC u_{2}\right)\sin\varphi}{R_{1}^{2}+R_{2}^{2}+2R_{1}R_{2}\cos\varphi}\,.\label{eq:vtot}
\end{equation}

The obtained total density current field $\VEC J_{\mathrm{tot}}(\VEC x,t)$
spanned by the various velocity components $\VEC v_{i}(\VEC x,t)$
and $\VEC u_{i\mathrm{R(L)}}(\VEC x,t)$ we have denoted as the ``path
excitation field'' \cite{Groessing.2012doubleslit}. It is built
by the sum of its partial currents, which themselves are built by
an amplitude weighted projection of the total current. Furthermore,
we observe that the superposition principle is violated for $\VEC J$,
and, analogously for $P,$ in the following sense: Due to the ``entanglement''
of partial probability densities with their corresponding partial
currents according to Eq.~(\ref{eq:Proj}), it generally holds that
\begin{equation}
\VEC J\mathrm{(}\VEC w_{\mathrm{i}}+\VEC w_{\mathrm{j}}\mathrm{)}\neq\VEC J\mathrm{(}\VEC w_{\mathrm{i}})+\VEC J\mathrm{(}\VEC w_{\mathrm{j}}),
\end{equation}
 since $\mathrm{(}\VEC w_{i}+\VEC w_{j}\mathrm{)}P(\VEC w_{i}+\VEC w_{j})$$\neq\VEC w_{i}P(\VEC w_{i})+\VEC w_{j}P(\VEC w_{j})$,
except for the special case of $P(\VEC w_{i}+\VEC w_{j})=\frac{P(\VEC w_{i})+P(\VEC w_{j})}{2}$,
which is fulfilled either by $P(\VEC w_{i})=P(\VEC w_{j})$ or $\VEC w_{i}=\VEC w_{j}$,
respectively. 

This result has to be interpreted in the following way. In orthodox
quantum mechanics the amplitudes of the wave function components have
to be summed up coherently in the case of undisturbed paths (``superposition''),
and for calculation of the probability density this sum has to be
taken as absolute value squared. Or, in other words, the Schrödinger
equation is linear, and observation of a state is regularized by Born's
rule. In our case, all the relevant variables, i.e.\ $P(\VEC w_{i})$
and $\VEC J\mathrm{(}\VEC w_{i})$ are not linear. Consequently, to
obtain the correct total probability density $P_{\mathrm{tot}}$ or
total current $\VEC J_{\mathrm{tot}}$, respectively, one has to take
into account \textit{all} elementary, i.e.\ partial contributions
to the corresponding variable.

Summarizing, the shift to a new representation for the emerging velocity
vectors (cf.\ Eq.~(\ref{eq:phi})) and the projection rule of Eq.~(\ref{eq:Proj})
build the kernel for a set of relations denoted as superclassical
current algebra. It is characterized by summing up the nonlinear,
probability-entangled partial currents, where each of the latter ones
contains information about the total field via a projection rule.
This property we have characterized as ``relational causality'':
Any change in a local field affects the total field, and \textit{vice
versa}. 

The trajectories or streamlines, respectively, are obtained according
to $\VEC{\dot{x}}=\VEC v_{\mathrm{tot}}$ in the usual way by integration.
Referring to \cite{Groessing.2012doubleslit}, we just mention that
by re-inserting the expressions for convective and diffusive velocities
$\VEC v_{i,\mathrm{conv}}=\frac{\nabla S_{i}}{m}$, $\VEC u_{i}=-\frac{\hbar}{m}$$\frac{\nabla R_{i}}{R_{i}}$,
one immediately identifies Eq.~(\ref{eq:vtot}) with the Bohmian
guiding equation and Eq.~(\ref{eq:Jfinal}) with the quantum mechanical
pendant for the probability density current \cite{Sanz.2008trajectory}. 

Again we have to emphasize that our result was achieved solely out
of kinematic relations by applying the rules mentioned above without
invoking complex $\psi$ functions or the like. However, as opposed
to the Bohmian theory, we obtain our results not in configuration
space but in common coordinate space. With respect to the following
discussion of the three-slit case, we can state that Eq.~(\ref{eq:Jfinal})
reflects the canonical result for the double slit in quantum mechanics,
i.e.\ the result for the probability density for detecting a particle
passing a double slit or a two-way interferometer undisturbed consists
of the joint probabilities of having only one slit open in each case
plus an interference term (in our case the sum of the sine and cosine
contributions). However, the validity of Born's rule will show only
after examination of the three-slit or three-beam case, respectively.

\section{Three-slit interference and Born's rule\label{sec:three}}

\begin{figure}[tbh]
\centering{}\includegraphics{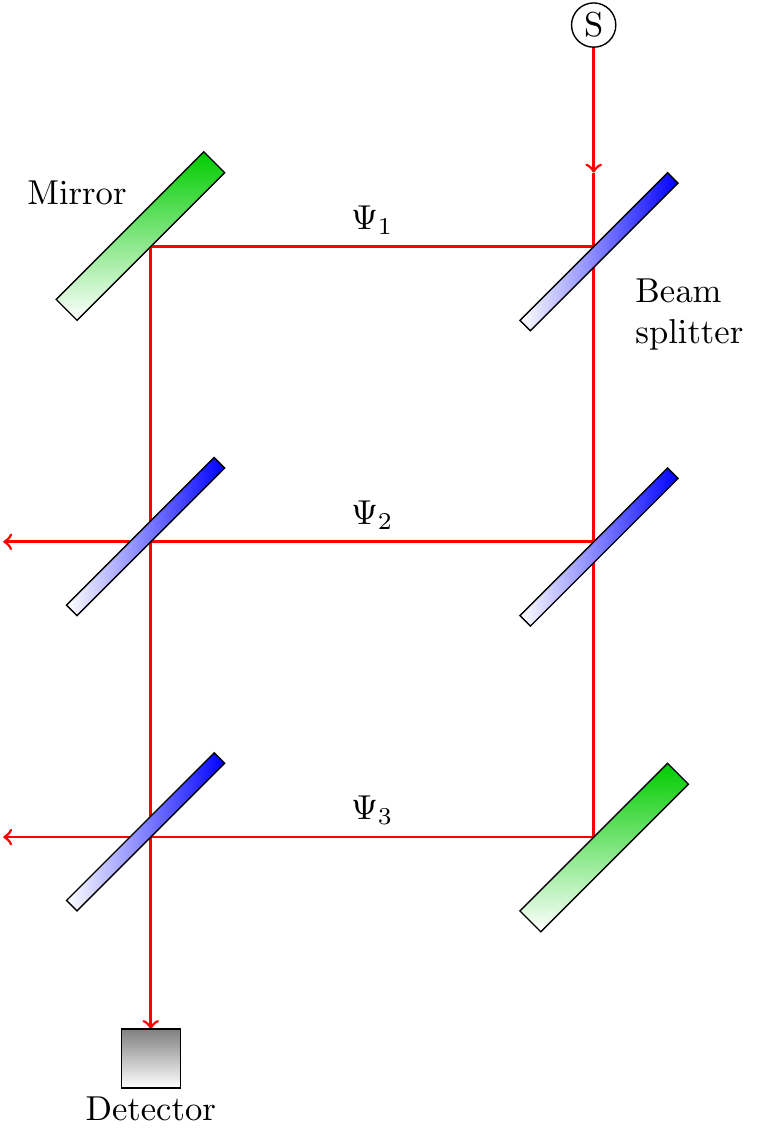}\caption{Schematic of a three path interferometer as analogue to the three
slit setup. \label{fig:three-path}}
\end{figure}

The extension to three slits, beams, or probability current channels,
respectively, is straightforward. Analogously to Fig.~\ref{fig:vektor}
we introduce a third emergent propagation velocity $\VEC v_{3}$ and
its corresponding diffusive velocities $\VEC u_{3\mathrm{L(R)}}$.
The phase shift of the third beam is denoted as $\chi$ and represents
the angle between the second and the third beam in our geometric representation
of the path excitation field. Analogously to the case of the double
slit, the three slit setup can be replaced by a three path interferometer
as shown in Fig.~\ref{fig:three-path}. According to Born's rule
the probability of even a single particle passing any of the three
slits splits into a sum of probabilities passing the slits pairwise,
i.e.\ going along both $\mathit{A}$ and $\mathit{B}$, $\mathit{B}$
and $\mathit{C,}$ or $\mathit{A}$ and $\mathit{C,}$ but never passing
$\mathit{A}$, $\mathit{B}$ and $\mathit{C}$ simultaneously. Whether
this splitting can be experimentally verified (cf.\ \cite{Sinha.2010ruling}
and \cite{Sollner.2012testing}) in exactly that way seems to be an
open question \cite{DeRaedt.2012analysis}, since it is argued that
with the decomposition of the three-slit wave function into its pairs
information about the original wave function is lost. Therefore, an
experimental test of Born's rule by measuring the outcome of blocked
pathways and summing them up seems not to be physically conclusive
\cite{DeRaedt.2012analysis}.

From the theoretical point of view interference-type phenomena have
been analyzed thoroughly \cite{Sorkin.1994quantum}, for the cases
of only one open slit up to $\mathit{N}$ open slits. For a double
slit setup the interference term is non-zero, i.e.\ $I_{AB}:=P_{AB}-P_{A}-P_{B}\neq0$,
with $P_{A(B)}$ being the detection probability with only one slit/path
$\mathit{A}$ or $\mathit{B}$, respectively, of a total of $\mathit{N}$
slits/paths open, and $P_{AB}$ for both slits $\mathit{A}$ and $\mathit{B}$
open. This well-known fact, representing the ``heart'' of quantum
mechanics, is to be contrasted with Sorkin's results for the following,
so-called ``second order sum rule'' \cite{Sorkin.1994quantum}:
\begin{align}
I_{ABC}:= & P_{ABC}-P_{AB}-P_{AC}-P_{BC}+P_{A}+P_{B}+P_{C}\\
= & P_{ABC}-(P_{A}+P_{B}+P_{C}+I_{AB}+I_{AC}+I_{BC})=0.\nonumber 
\end{align}
This result is remarkable insofar as it can be inferred that interference
terms theoretically always originate from pairings of paths, but never
from triples etc. Any violation of this second order sum rule, i.e.\ $I_{ABC}$$\neq0$,
and thus of Born's rule would have dramatic consequences for quantum
theory like a modification of the Schrödinger equation, for example.

By returning to our model, the total probability density current for
three paths is calculated according to the rules set up in Section~\ref{sec:current}.
We adopt the notations of the two slit system also for three slits,
i.e.\ now employing nine velocity contributions: $\VEC v_{i}$, $\VEC u_{i\mathrm{R(L)}}$,
$i=1,2,3$. Analogously, the three generally different amplitudes
are denoted as $\mathrm{R}\mathrm{(}\VEC v_{i})=\mathrm{R}\mathrm{(}\VEC u_{i\mathrm{R}})=\mathrm{R}\mathrm{(}\VEC u_{i\mathrm{L}})=R_{i}$,
$i=1,2,3$. We keep the definition of $\varphi$ as $\varphi:=\arccos(\VEC{\hat{v}}_{1}$$\cdot$$\VEC{\hat{v}}_{2})$,
and we define the second angle as $\chi:=\arccos(\VEC{\hat{v}}_{2}\cdot\VEC{\hat{v}}_{3}).$
Similarly to Eq.~(\ref{eq:triv}), the diffusive velocities $\VEC u_{i\mathrm{\mathrm{R}}}-\VEC u_{i\mathrm{L}}$
combine to $\VEC u_{i}$, $i=1,2,3$, thus ending up with $2N=6$
effective velocities. Therefore we obtain, analogously to the calculation
in the previous section, 
\begin{align}
\VEC{J_{\mathrm{tot}}=} & R_{1}^{2}\VEC v_{\mathrm{1}}+R_{2}^{2}\VEC v_{\mathrm{2}}+R_{3}^{2}\VEC v_{\mathrm{3}}+R_{1}R_{2}\left(\VEC v_{\mathrm{1}}+\VEC v_{\mathrm{2}}\right)\cos\varphi+R_{1}R_{2}\left(\VEC u_{\mathrm{1}}-\VEC u_{\mathrm{2}}\right)\sin\varphi\nonumber \\
 & +R_{1}R_{3}\left(\VEC v_{\mathrm{1}}+\VEC v_{\mathrm{3}}\right)\cos\left(\varphi+\chi\right)+R_{1}R_{3}\left(\VEC u_{\mathrm{1}}-\VEC u_{\mathrm{3}}\right)\sin\left(\varphi+\chi\right)\nonumber \\
 & +R_{2}R_{3}\left(\VEC v_{\mathrm{2}}+\VEC v_{\mathrm{3}}\right)\cos\chi+R_{2}R_{3}\left(\VEC u_{\mathrm{2}}-\VEC u_{\mathrm{3}}\right)\sin\chi\label{eq:3.1}
\end{align}
and
\begin{align}
P_{\mathrm{tot}} & =R_{1}^{2}+R_{2}^{2}+R_{3}^{2}+2R_{1}R_{2}\cos\varphi+2R_{1}R_{3}\cos\left(\varphi+\chi\right)+2R_{2}R_{3}\cos\chi\label{eq:p3}\\
 & =P(\VEC v_{\mathrm{1}})+P(\VEC v_{\mathrm{2}})+P(\VEC v_{3}).\nonumber 
\end{align}

In analogy to the double slit case (cf.\ Eq.~(\ref{eq:Ptot2slit}))
we obtain -- at first sight -- a \textit{classical} Kolmogorov sum
rule for the probabilities on the one hand, but also the complete
interference effects for the double, three- and, as will be shown
in the next section, for the $N$-slit cases, on the other hand. However,
the particular probabilities $P(\VEC v_{i})$ in Eq.~(\ref{eq:Ptot2slit})
and Eq.~(\ref{eq:p3}), do not correspond to the probabilities of
the assigned slits if solely opened, i.e.\ $P_{AB}(\VEC v_{\mathrm{1}})=\left(R_{1}^{2}+R_{1}R_{2}\cos\varphi\right)\neq P_{A}(\VEC v_{\mathrm{1}})=R_{1}^{2}$.
Consequently, each of the probability summands in said equations does
\textit{not} correspond to an independent probability of the respective
slit if solely opened. Note that our result reflects an illustrative
remark of Ballentine \cite{Ballentine.1986probability} stating the
fact that $I_{AB}\neq0$ for the double slit experiment does \textit{not}
mean that the classical probability sum rules are violated, since
they are originally formulated for \textit{mutually exclusive} states.
By keeping in mind that said probabilities $P_{A}$, $P_{B}$, and
$P_{AB}$ are in fact \textit{conditional} probabilities, there is
no violation of any classical probability sum rule by stating the
experimental observation $P_{AB}\neq P_{A}+P_{B.}$ Translated into
our double slit model, we have in the case of both slits $\mathit{A}$
and $\mathit{B}$ open, $P_{AB}=:P_{\mathrm{tot}}(A\wedge B)=P(\VEC v_{\mathrm{1}})+P(\VEC v_{\mathrm{2}})=R_{1}^{2}+R_{2}^{2}+2R_{1}R_{2}\cos\varphi$,
which clearly must not be confused with the mutually exclusive case
$P_{\mathrm{tot}}(A\vee B)=P_{A}(\VEC v_{\mathrm{1}})+P_{B}(\VEC v_{\mathrm{2}})=R_{1}^{2}+R_{2}^{2}.$

Finally, we obtain for the cases of one (i.e.\ $N=\mathit{A}$),
two and three open slits, respectively, 
\begin{equation}
I_{A}=P_{A}(\VEC v_{\mathrm{1}})=R_{1}^{2},
\end{equation}
\begin{equation}
I_{AB}=P_{AB}-P_{A}(\VEC v_{\mathrm{1}})-P_{B}(\VEC v_{\mathrm{2}})=2R_{1}R_{2}\cos\varphi,\label{eq:1storder}
\end{equation}
\begin{equation}
I_{ABC}=P_{ABC}-P_{AB}-P_{AC}-P_{BC}+P_{A}(\VEC v_{\mathrm{1}})+P_{B}(\VEC v_{\mathrm{2}})+P_{C}(\VEC v_{3})=0\,,\label{eq:2ndorder}
\end{equation}
where $P_{AB}$ is assigned to $P_{\mathrm{tot}}$ of Eq.~(\ref{eq:Ptot2slit})
and $P_{ABC}$ to $P_{\mathrm{tot}}$ of Eq.~(\ref{eq:p3}). In the
double slit case, e.g., with slits $\mathit{A}$ and $\mathit{B}$
open, we obtain the results of (\ref{eq:Ptot2slit}). If $\mathit{B}$
were closed and $\mathit{C}$ were open instead, we would get the
analogous result, i.e.\ $\VEC v_{\mathrm{2}}$ and $\varphi$ replaced
by $\VEC v_{\mathrm{3}}$ and $\varphi_{1,3}$. If all three slits
$\mathit{A,B},C$ are open, we can use the pairwise permutations of
the double slit case, i.e.\ $\mathit{A}\wedge B$, $A\wedge C$,
or $\mathit{B\wedge C}$, respectively, with $\varphi_{1,3}$ identified
with $\left(\varphi+\chi\right)$, etc. Thus we conclude that in our
model the addition of ``sub-probabilities'' indeed works and provides
the correct results.

Summarizing, with our superclassical model emerging out of a sub-quantum
scenario we arrive at the same results as standard quantum mechanics
fulfilling Sorkin's sum rules \cite{Sorkin.1994quantum}. Opposed
to the open question in quantum mechanics of whether said decomposition
of a three-slit probability term into its sum of double- and one-slit
probabilities only represents a ``mathematical trick'' \cite{DeRaedt.2012analysis},
we observe the following: Whereas in standard quantum mechanics Born's
rule originates from building the squared absolute values of additive
$\psi$ functions representing the probability amplitudes for different
paths, in our case we obtain the pairing of paths as a natural consequence
of \textit{pairwise} selection of unit vectors of all existing velocity
components constituting the probability currents. Thus we obtain \textit{all}
possible pathways within an $\mathit{N}$-slit setup by a two-channel
projection method. The sum rules, Eqs.~(\ref{eq:Proj}) through (\ref{eq:vtot_fin}),
guarantee that each partial contribution, be it from the velocity
contributions within a particular channel or from different channels,
accounts for the final total current density for each point between
source and detector. Since for only one slit open the projection rule
(\ref{eq:Proj}) trivially leads to a linear relation between $\mathit{P}$
and $R^{2}$, the asymmetry between the latter quantities, due to
the nonlinear projection rule, becomes effective for $N\geq2$ slits
open. Consequently, the violation of the first order sum rule (\ref{eq:1storder}),
i.e.\ $I_{AB}\neq0$, represents a \textit{natural} result of our
principle of relational causality. Moreover, as we have argued above,
the opening of an additional slit solely adds pairwise path combinations.
As all higher interference terms have already incorporated said asymmetry,
the result can finally be reduced to the double slit case, thus yielding
a zero result as in (\ref{eq:2ndorder}) according to Sorkin's analysis.

This is a further hint that our model can reproduce all phenomena
of standard quantum theory with the option of giving a deeper reasoning
to principles like Born's rule or the hierarchical sum-rules, respectively,
with the prospect of a physics beyond quantum theory (cf.\ the discussion
in the last section).

\section{$N$-slit interference and the quantum Talbot effect\label{sec:nslit}}

We can already infer from the three-slit device that due to the pairwise
selection of the velocity field components $\VEC v_{i}(\VEC x,t)$
and $\VEC u_{i\mathrm{L(R)}}(\VEC x,t)$, $i=1,\ldots,N$, the interference
effect of every higher order grating can be reduced to successive
double-slit algorithms. For a compact description of the $\mathit{N}$-slit
case we return to the notation (\ref{eq:nslit.2.4}) and (\ref{eq:nslit.2.5})
of general velocity vectors $\VEC w_{i}$ with
\begin{equation}
\VEC w_{1}:=\VEC v_{\mathrm{1}},\quad\VEC w_{2}:=\VEC u_{\mathrm{1R}},\quad\VEC w_{3}:=\VEC u_{\mathrm{1L}},\quad\VEC w_{4}:=\VEC v_{\mathrm{2}},\ldots,\quad\VEC w_{3N}:=\VEC u_{N\mathrm{L}},
\end{equation}
with $\VEC w_{3i-2}:=\VEC v_{i}$ referring to the propagation velocities,
$\VEC w_{3i-1}:=\VEC u_{i\mathrm{R}}$ and $\VEC w_{3i}:=\VEC u_{i\mathrm{L},}$
with $i=1,\ldots,N$ denoting the diffusion velocities for each channel
$\mathit{i.}$ According to the Eqs.~(\ref{eq:Proj}) to (\ref{eq:Jtot6}),
now with a general number $\mathit{N}$ of slits, the calculation
for the total probability density is straightforward, since only the
contributions of the propagation velocities are non-zero:
\begin{align}
P_{\mathrm{tot}}(N) & =\left(\vphantom{\overset{{\scriptstyle N}}{\underset{{\scriptstyle i=1}}{\sum}}}R(\VEC w_{1})\VEC{\hat{w}}_{1}+R(\VEC{w_{\mathrm{4}}})\VEC{\hat{w}}_{4}+...+R(\VEC w_{3N-2})\VEC{\hat{w}}_{3N-2}\right)^{2}=\left(\overset{{\scriptstyle N}}{\underset{{\scriptstyle i=1}}{\sum}}R(\VEC w_{3i-2})\VEC{\hat{w}}_{3i-2}\right)^{2}\nonumber \\
 & =\overset{}{\overset{{\scriptstyle N}}{\underset{{\scriptstyle i=1}}{\sum}}\left(R^{2}(\VEC w_{3i-2})+\sum_{j=i+1}^{N}2R(\VEC w_{3i-2})R(\VEC w_{3j-2})\cos\varphi_{i,j}\right)}.
\end{align}
For the current density we can generalize the relations (\ref{eq:triv})
to $\VEC{\hat{w}}_{3i-1}\mathrm{\mathit{R\mathrm{(}\VEC w_{3i-1}\mathrm{)}}}+\VEC{\hat{w}}_{3i}\mathrm{\mathit{R\mathrm{(}\VEC w_{3i}\mathrm{)}}}=0$,
and obtain 
\begin{equation}
\VEC J_{\mathrm{tot}}=\sum_{i=1}^{3N}\VEC J(\VEC w_{i})=\sum_{i=1}^{3N}\left(R(\VEC w_{i})\VEC w_{i}{\displaystyle \cdot\sum_{j=1}^{N}}\VEC{\hat{w}}_{3i-2}R(\VEC w_{3j-2})\right).
\end{equation}
The partial current summands for the propagation velocities $\VEC w_{3i-1}$,
$i=1,\ldots,N$, read as
\begin{equation}
\VEC J(\VEC w_{3i-2})=\VEC w_{3i-2}\left(R^{2}(\VEC w_{3i-2})+\sum_{j\neq i=1}^{N}R(\VEC w_{3i-2})R(\VEC w_{3j-2})\cos\varphi_{i,j}\right).
\end{equation}

Since the calculations for the diffusive currents are lengthy, but
straightforward, we now present the final results. We use our previous
notation for the final $2N$ effective velocity contributions with
$\mathit{N}$ slits 
\begin{align}
\VEC w_{3i-2}:=\VEC v_{i},\;\VEC w_{3i-1}-\VEC w_{3i}:=\VEC u_{i},\; R(\VEC w_{3i-2})=R(\VEC w_{3i-1})=R(\VEC w_{3i})=:R_{i},\quad i=1,\ldots,N,
\end{align}
leading to
\begin{align}
P_{\mathrm{tot}}(N) & =\left(\sum_{i=1}^{N}R_{i}\VEC{\hat{v}}_{i}\right)^{2}={\displaystyle \sum_{i=1}^{N}}P(\VEC v_{i})=\sum_{i=1}^{N}\left(R_{i}^{2}+\sum_{j=i+1}^{N}2R_{i}R_{j}\cos\varphi_{i,j}\right),\\
\nonumber \\
\VEC J_{\mathrm{tot}}(N) & =\sum_{i=1}^{N}\left(R_{i}^{2}\VEC v_{i}+\overset{}{\sum_{j=i+1}^{N}R_{i}R_{j}\left\{ \vphantom{\sum_{i=1}^{N}}\left(\VEC v_{i}+\VEC v_{j}\right)\cos\varphi_{i,j}+\left(\VEC u_{i}-\VEC u_{j}\right)\sin\varphi_{i,j}\right\} }\right)\,.
\end{align}

From these results we can clearly see that the addition of an arbitrary
number of slits represents a simple inductive extension from the double
slit case as we had stated in the previous section.

\begin{figure}[!tb]
\begin{centering}
\includegraphics[scale=0.95]{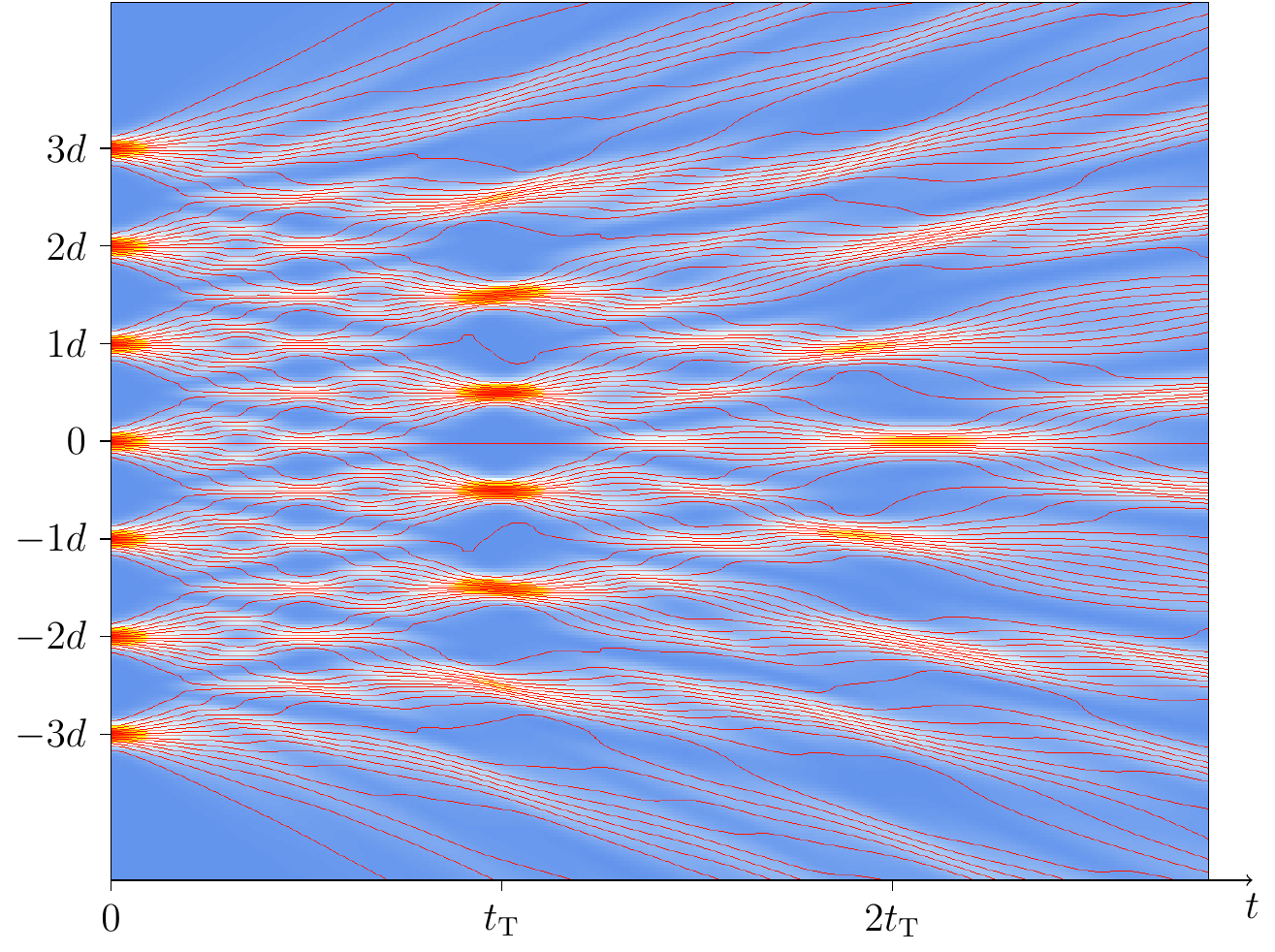}\vspace*{4mm}

\par\end{centering}

\centering{}\includegraphics[scale=0.95]{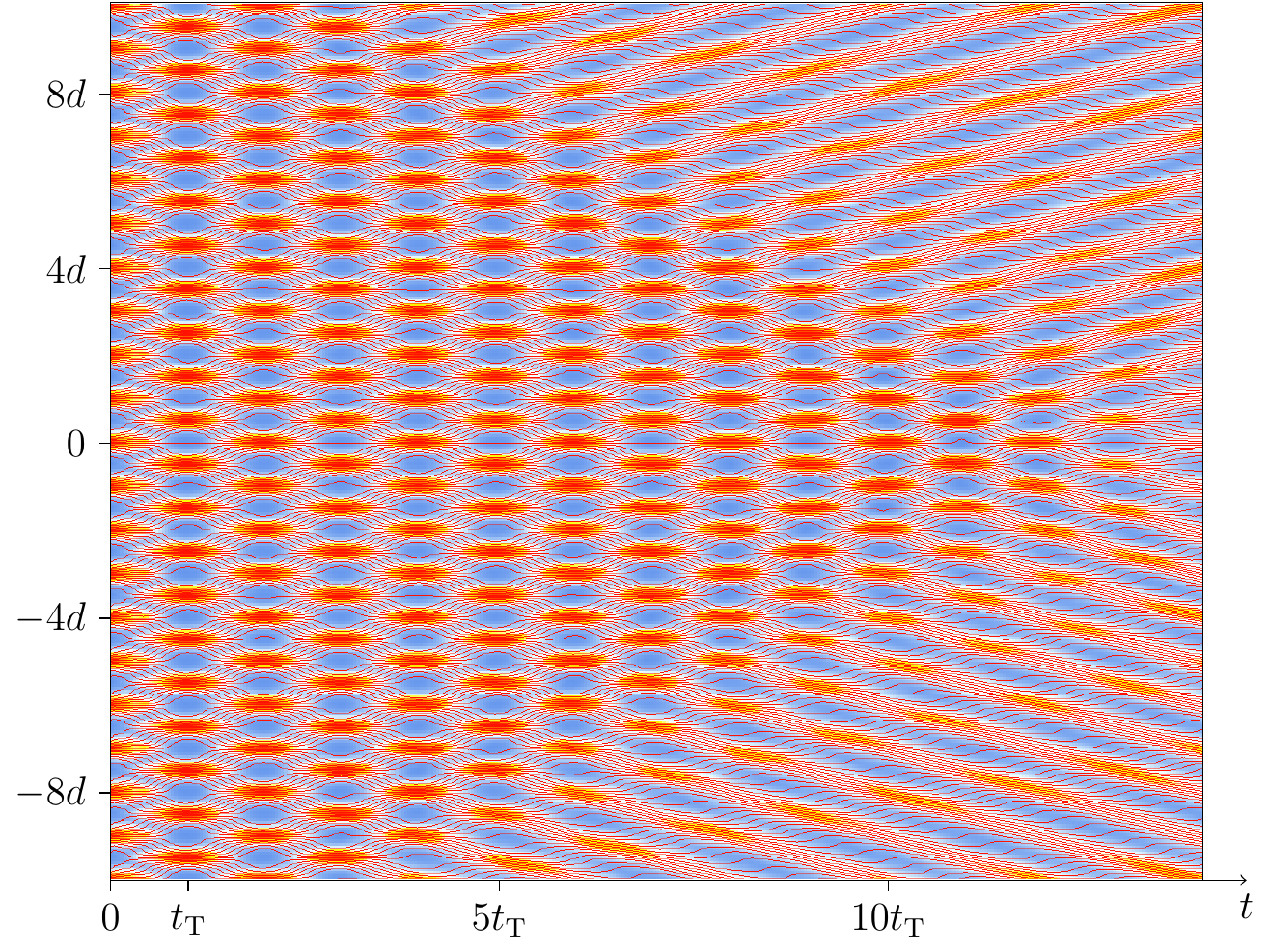}\caption{Intensity distributions via classical computer simulations of the
Talbot carpet for a 7-slit ($d=1.06\,\mathrm{nm}$) and a 27-slit
($d=0.53\,\mathrm{nm}$) setup of Table~\ref{tab:TalbotCarpetParameters},
respectively. Averaged particle trajectories are displayed in red.
\label{fig:Born.5}}
\end{figure}

In well-known manner one obtains the trajectories from $\VEC{\dot{x}}_{\mathrm{tot}}=\VEC v_{\mathrm{tot}}=\frac{\VEC J_{\mathrm{tot}}}{P_{\mathrm{tot}}}$
\cite{Sanz.2008trajectory}. As opposed to this analytical procedure,
we use the simulation tools disclosed in \cite{Mesa.2013variable},
which are displayed in the computer simulations of Fig.~\ref{fig:Born.5}
for a 7-slit and a 27-slit setup, respectively. Already for the 7-slit
case one can observe the emergence of a repetitive short range pattern
until the Fraunhofer regime is reached. At the so-called Talbot distance
\begin{equation}
z_{\mathrm{T}}=d^{2}/\lambda,\label{eq:talbot}
\end{equation}
where $\mathit{d}$ denotes the grating period and $\lambda$ the
wavelength of the incident plane wave, the initial patterns of the
7 vertically arranged slit openings reappear with a shift of $d/2.$
Table~\ref{tab:TalbotCarpetParameters} shows the results for different
values of $\lambda$ and $d$ and compares them with the observed
values $y_{\mathrm{T}}$ of the Talbot distance for the 7- and various
$N$-slit cases. According to Table~\ref{tab:TalbotCarpetParameters},
we use the parameters for neutrons as 
\begin{table}[!tb]
\begin{centering}
\renewcommand*\arraystretch{1.5}%
\begin{tabular}{|c|c|c|c|c|}
\hline 
Setup & a & b & c & d\tabularnewline
\hline 
\hline 
$\lambda$ & 1 nm & 1 nm & 1 nm & 1 nm\tabularnewline
\hline 
$d$ & 0.53 nm & 1.06 nm & 1.59 nm & 2.12 nm\tabularnewline
\hline 
$z_{\mathrm{T}}$ & 0.28 nm & 1.13 nm & 2.53 nm & 4.5 nm\tabularnewline
\hline 
$y_{\mathrm{T,7-slit}}$ & 0.28 nm & 1.14 nm & 2.53 nm & 4.52 nm\tabularnewline
\hline 
$y_{\mathrm{T,}N\mathrm{-slit}}$ & 0.29 nm ($N=27$) & 1.13 nm ($N=27)$ & 2.53 nm ($N=25$) & 4.49 nm ($N=13)$\tabularnewline
\hline 
\end{tabular}
\par\end{centering}

\caption{Parameters for the Talbot carpet simulations of Fig.~\ref{fig:Born.5}.
\label{tab:TalbotCarpetParameters}}
\end{table}
$d=1.06\,\mathrm{nm}$ and $\lambda=1\,\mathrm{nm}$, with a neutron
mass $m_{\mathrm{n}}=1.675\cdot10^{-27}\,\mathrm{kg}$. The spatial
step width is chosen as $\Delta x=0.0378\,\mathrm{nm}$, the time
resolution as $\Delta t=1.92\cdot10^{-14}\,\mathrm{s}$. Said shifted
reappearance of the pattern occurs for the first time at time step
150, i.e.\ at $t_{\mathrm{T}}=150\cdot\Delta t=2.88\cdot10^{-12}\,\mathrm{s}$.
The standard transformation to the two-dimensional case \cite{Holland.1993}
by re-parametrizing the $t$-axis via $y=\hbar k_{\mathrm{n}}\Delta t/m_{\mathrm{n}}=h\Delta t/(\lambda m_{\mathrm{n}})$
leads to the observed distance $y_{\mathrm{T}}=ht_{\mathrm{T}}/(\lambda m_{\mathrm{n}})=1.14\,\mathrm{nm}$,
which matches nicely with the formula of the Talbot distance $z_{\mathrm{T}}$
of (\ref{eq:talbot}). The observed values for the Talbot distance
$y_{\mathrm{T}}$ in our discretized model agree for any $N$-slit
setup as expected in accordance with Eq.~(\ref{eq:talbot}), which
only depends on $d$ and $\lambda$. Moreover, we also obtain the
correct results for any other choice of $m$ or $\lambda$.

For multiples of $2z_{\mathrm{T}}$ the recurrence of the original
state is observed, as it is particularly obvious in the case of 27
slits. Due to the non-crossing of all trajectories, as it has been
discussed in \cite{Groessing.2012doubleslit}, the ``caverns'' in
the middle stay confined until they are broken up by the influence
of the boundary area via the ``light cone''. In the limit of an
indefinitely extended grating the pattern clearly would be maintained
\textit{ad infinitum}. 

Since the averaged trajectories obtained with our superclassical current
algebra are identified with the Bohmian trajectories of Sanz \textit{et\,al.}~\cite{Sanz.2007causal},
we have thus shown that the emerging quantum carpet for $\mathit{N}$
slits constituted by characteristic repetitive patterns can be reproduced
without any (real or complex) quantum mechanical state function.

\section{Conclusions and outlook\label{sec:conclusion}}

It has been shown in a series of papers \cite{Groessing.2012doubleslit,Groessing.2008vacuum,Groessing.2009origin,Groessing.2010emergence,Groessing.2011dice,Groessing.2011explan,Groessing.2013dice}
that phenomena of standard quantum mechanics like Gaussian dispersion
of wave packets, superposition, double slit interference, Planck's
energy relation, or the Schrödinger equation, can be assessed as the
emergent property of an underlying sub-structure of the vacuum combined
with diffusion processes reflecting also the stochastic parts of the
zero-point field, i.e.\ the zero-point fluctuations \cite{Cetto.2012quantization}.
Thus we obtain the quantum mechanical results as an averaged behavior
of sub-quantum processes. The inclusion of relativistic physics has
not been considered yet, but should be possible in principle. 

In the present paper we have started with a ``minimal set'' of assumptions
and results from our previous work, like the use of classical propagation
velocities $\VEC v_{i}(\VEC x,t)$ and diffusive velocities $\VEC u_{i}(\VEC x,t)$,
the orthogonality relation between them and the inclusion of the phase
angle between the propagation velocity vectors of the emerging path
excitation field, the latter one spanned by said velocity vectors
and additionally weighted with the corresponding probability densities. 

We introduced an amplitude weighted projection rule to account for
all partial velocity contributions to a partial current $\VEC J(\VEC w_{i})$
per velocity channel, whereby the symmetry between amplitude and the
probability density, usually given by the squared amplitude, is broken.
The total wave intensity field accounting for the emerging thermal
landscape consists of the sum of all \textquoteleft{}local\textquoteleft{}
intensities in each channel, and the \textquoteleft{}local\textquoteleft{}
intensity in each channel is the result of the interference with the
total intensity field. The mutual dependence of a total current and
its parts we denoted as ``relational causality'', and this represents
an essential part of the calculus which we subsumed as superclassical
current algebra, combining the physics of different scales, e.g.,
sub-quantum and classical macro physics. The presence of \textit{all}
velocity channels at any point of the spatio-temporal domain between
source and detector is based upon our ansatz that $\mathit{N}$ Gaussians
(representing the wave amplitudes of a particle immediately after
passing $\mathit{N}$ slits/paths) do not have any artificial cut-off,
but actually extend across the whole slit/path system \cite{Groessing.2012vaxjo,Groessing.2013dice}.
This is supported by experimental evidence showing that interference
can be caused by the nonlocally far-reaching action of the plane-waves
of a quantum mechanical wave-function \cite{Rauch.2006hidden,Rauch.2012particle},
and by the works of Mandelis \cite{Mandelis.2000diffusion,Mandelis.2001diffusion-wave},
where diffusion wave fields are related to oscillating sources extending
nonlocally across the whole domain of an experimental setup.

As an important result, and as a natural consequence of our considerations,
a third order interference term violating Born's rule in orthodox
quantum mechanics is absent in our superclassical framework. We have
shown, on the one hand, that the step from a one-slit/path setup to
the double slit/path case introduces a new quality due to the nonlinear
projection rules, which is kept in all higher extensions. On the other
hand, the total probability for the $\mathit{N}$-slit/path case is
built by the summation of ``sub-probabilities'', i.e.\ the probabilities
for ($\mathit{N-1}$)-slit/path configurations. Altogether, the double
slit/path case represents an exceptional system insofar as a new quality
interference appears and all higher ($N>2$) configurations are reducible
to it. Consequently, the pairwise path selection, the violation of
the first order sum rule, and the validity of all higher order sum
rules are explained naturally within our theoretical framework.

Furthermore, we have derived general formulas for the $N$-slit current
densities and thus are able to give a micro-causal account for the
kinematics of the quantum Talbot effect. The Talbot distance can be
reproduced also quantitatively in our model.

Throughout the whole paper we have made use of \textit{averaged} diffusive
velocities emerging from billions of billions sub-quantum fluctuations.
Therefore, within the framework of our theory we can tackle questions
going beyond standard quantum theory. At the emerging quantum level,
i.e.\ at times $t\ggg1/\omega$, with $\omega$ representing the
familiar zitterbewegung frequency, e.g., for the electron $\omega\approx10^{21}\,\mathrm{Hz}$,
we obtain exact results strongly suggesting the validity of Born's
rule, for example. However, approaching said sub-quantum regions by
increasing the time resolution to the order of $t\approx1/\omega$
suggests a possibly gradual breakdown of said rule, since the averaging
of the diffusive and convective velocities and their mutual orthogonality
of the averaged velocities is not reliable any more. In principle,
this should eventually be testable in experiment.

\section*{Acknowledgments}

We thank Hans De Raedt for pointing out ref.~\cite{Ballentine.1986probability}
to us, Jan Walleczek for many enlightening discussions, and the Fetzer
Franklin Fund for partial support of the current work.

\providecommand{\href}[2]{#2}\begingroup\raggedright\endgroup

\end{document}